\documentclass{ws-p8-50x6-00}

\newcommand{\chiPT}{$\chi{\rm PT}$}
	
\begin{document}

\title{Baryon Spectroscopy on the Lattice}

\author{Robert G. Edwards}

\address{Jefferson Lab\\
12000 Jefferson Avenue
Newport News, Virginia 23606, USA\\ 
E-mail: edwards@jlab.org}

\maketitle

\abstracts{
Recent lattice QCD calculations of the baryon spectrum are outlined.
}

\section{Introduction}

Quantum Chromodynamics (QCD) provides an excellent description of
nature; however, the theory suffers from divergences that must be
removed to render it finite. Lattice QCD provides an {\em apriori}
non-perturbative regularization of QCD that makes it amenable to
analytic and computational methods. No model assumptions other than
QCD itself are needed to formulate the theory. This review surveys the
rapidly evolving work in using Lattice QCD for calculations of
baryon spectroscopy.  Along the way, sources of systematic
uncertainties in calculations are described and future directions are
outlined.

\subsection{Regularization of QCD on a lattice}

As the starting point for lattice QCD, the path integral formulation
in Euclidean space is used\cite{Montvay_book}.  The usual continuous
space-time of 4-dimensional continuum QCD are approximated with a
discrete 4-dimensional lattice, with derivatives approximated by
finite differences.  Quarks are put on sites, gluons on links. Gluons
are represented as $3\times 3$ complex unitary matrices $U_\mu(x) =
\exp(iga A_\mu(x))$ elements of the group SU(3) with vector potential
$A_\mu(x)$, coupling $g$, and lattice spacing $a$. The vacuum
expectation value of operators involves path integration over gauge
and fermion fields
\[
\displaylines{
\left\langle {O\left( {U,\psi ,\overline \psi  } \right)} \right\rangle  = 
 \frac{1}{Z}\int {dU_\mu  d\overline \psi d\psi\,O\left({U,\psi,\overline \psi}\right)\,\,
e^{-S_G\left( U \right) + \overline \psi  M\left( U \right)\psi } }  \cr 
   \rightarrow \frac{1}{Z'}\int {dU_\mu \,O\left({U,M^{-1}\left(U\right)}\right)\,\,\det 
  \left({M\left(U \right)} \right)e^{-S_G \left(U \right)}}\quad. \cr}
\]
\noindent
The Gaussian integration over the anti-commuting fermion fields $\psi$ resulted
in the ${\rm det}(M(U))$ and $M^{-1}(U)$ factors with $M(U)$ a lattice form of the
Dirac operator. The gauge action $S_G(U)$ approximates the Yang-Mills action
of the continuum. The quenched approximation neglects the fermion determinant.

The choice of working in Euclidean space resulted in no factors of $i$
in the exponent multiplying the gauge and fermion actions. The path
integral therefore resembles a 4-dimensional statistical mechanical model making
it amenable to analytic as well as Monte Carlo methods for evaluation.


Numerical predictions from lattice QCD are in principle exact (to some
precision) after systematic errors are controlled.  The statistical
uncertainties go like $1/\sqrt{N}$ for N configurations of gauge
fields in a Monte Carlo ensemble average. Systematic uncertainties
include: (1) Finite volume - the lattice box must hold a hadron state,
typically a lattice size of $L \sim 2$fm or more is needed. Several
pion Compton wavelengths are needed $m_\pi L \sim 4$. (2) Chiral
extrapolations - calculations with small quark masses are expensive -
extrapolate observables to physical quark mass region (delicate!).
(3) Discretization effects: inherent ${\cal O}(a)$ or ${\cal O}(a^2)$
lattice uncertainty. One must extrapolate to continuum limit $(a
\rightarrow 0)$ to recover physical quantities.

\section{Confinement and Model Predictions - Static Quark Potentials }



A particularly useful r\^ole of lattice QCD is model testing.  There
is significant recent activity in the study of 3 quark potentials
which provide phenomenological insight into the forces inside a
baryon. By gauge invariance, the quarks must be joined by 3 glue
strings. These strings meet at a ``gluon junction'', which has been
conjectured to be a non-perturbative excitation of the QCD
vacuum\cite{Kharzeev_96}.  What is the area law behavior? One can test
two ans\"atze.

The {\em Y}-ansatz predicts the potential grows linearly like $V_{qqq}\propto
\sigma_{qq} L_Y$ where $L_Y$ is the minimal length of the 3 flux tubes
necessary to join the 3 quarks at the Steiner point. It is derived
from strong coupling arguments\cite{Isgur_paton}, and is consistent
with the dual superconductivity confinement scenario.

At large distances, the $\Delta$-ansatz predicts instead that the
potential grows linearly with the perimeter $L_\Delta$ of the quark
triangle, e.g.  $V_{qqq}\propto \sigma_{q\bar{q}} L_\Delta/2$. It is
derived from a model of confinement by center vortices using a
topological argument\cite{Cornwal_XX}.
%

There is controversy as to which
ansatz\cite{Takahashi_02,Alexandrou_02} holds. Recent
work\cite{Alexandrou_02} claims that at short distances the potential
approaches the $\Delta$-ansatz but rises like the {\em Y}-ansatz at
large distances.  Departures from the $\Delta$-ansatz appear above
$d_{qq}\sim 0.7$fm hence the $\Delta$ model is more appropriate inside
a hadron. However, recently Simonov\cite{Simonov_02} claims there is
a field strength depletion near the {\em Y}-junction which lowers the
potential and could disguise the true behavior. Tests using 
adjoint sources could help reconcile the various claims.





\section{Hadron Spectrum }
\subsection{Chiral Symmetry}
As mentioned before, for accurate lattice calculations
systematic uncertainties need to be controlled.
The discretization of the Dirac operator has been particularly
troublesome since lattice QCD's inception and can significantly affect
continuum and chiral extrapolations. The ``doubling'' problem
is easily demonstrated by examining the lattice momentum representation
of the free Dirac operator, namely $\sum_\mu {\gamma_\mu
\partial_\mu} \rightarrow \frac{i}{a}\sum_\mu {\gamma_\mu
\sin(a p_\mu)}$.  The propagator has additional zeros at the momentum
corners, e.g. $a p_\mu=0,\pi$ so there are 16 species of fermions in
general. Originally, Wilson lifted the doublers by adding a Laplacian
term that breaks chiral symmetry.  In fact, the Nielson-Ninomia no-go
theorems state one cannot avoid both doubling and chiral symmetry
breaking with a local, hermitian action analytic in the gauge fields.
This major theoretical problem has been solved with the recent advent
of chiral fermion actions\cite{Chiral} (e.g., Domain-Wall or
Overlap fermions) and their use is crucial for matrix elements. How important is
chiral symmetry for spectroscopy studies?

Renormalization theory tells us that breaking a symmetry leads to
induced quantum terms in an action. The Wilson fermion action has
${\cal O}(a)$ scaling from the breaking of chiral symmetry. One can
add a dimension 5 operator (hyper-fine term) and rigorously improve
scaling from ${\cal O}(a)$ to ${\cal O}(a^2)$. Scaling violations are
dramatically reduced -- mostly from improving chiral symmetry. Scaling
violations are comparable with chiral fermion formulations.  The
conclusion is that chiral symmetry is important for accurate spectrum
calculations\cite{Edwards_98} at comparatively heavy quark masses.
However, the benefits of chiral fermion actions with {\em exact}
chiral symmetry are now being dramatically demonstrated as near
physical quark masses are approached as will be shown below.


\subsection{Quenched Pathologies in Hadron Spectrum}
Clearly, precisely controlled lattice calculations come with the
inclusion of the fermion determinant. However, because of their vastly
reduced computational cost quenched calculations are quite prevalent
and one can gain important phenomenological insight into QCD, but the
potentially large systematic errors induced in this approximation
should be carefully ascertained.  Suppressing the fermion determinant
leads to well known pathologies as studied in chiral perturbation
theory\cite{chipt}.  There are missing vacuum contributions to the
disconnected piece of singlet correlators
\[
\displaylines{
  \left\langle {\overline \psi(x)\gamma_5 \psi(x)
     \overline \psi(y)\gamma_5 \psi(y)} \right\rangle \, = 
  \,\left\langle {{\rm Tr}\left[ {\gamma_5 G\left( {x,y} \right)\,
  \gamma_5 G\left( {y,x} \right)} \right]_{c,s} } \right\rangle  \cr 
   - \,N_f \left\langle {{\rm Tr}\left[{\gamma _5 G\left( {x,x} \right)\,} 
   \right]_{c,s} {\rm Tr}\left[ {\,\gamma _5 G\left( {y,y} \right)}\right]_{c,s} } 
     \right\rangle \quad . \cr}
\]
These effects are manifested in the $\eta'$ propagator missing vacuum contributions
with new double pole divergences arising of the form 
\begin{equation}
\int \frac{d^4 p}{(2\pi)^4} \; e^{i p\cdot x}\,
\langle {\rm Tr}\gamma_5 G(x,x){\rm Tr}\gamma_5 G(0,0)\rangle =
f_P\frac{1}{p^2+m^2_\pi} m_0^2\frac{1}{p^2+m^2_\pi} f_p \;+\;\ldots
\label{eq:mass_insertion}
\end{equation}
How dramatic are these quenched effects and to what extent do they affect the
extraction of physical observables? One idea is to incorporate knowledge of 
quenched divergences in calculations and then attempt to extract useful information.

\subsection{Decay in the Quenched Approximation}
\label{sec:decay}

The $m_0^2$ in Eq.~(\ref{eq:mass_insertion}) is the mass shift needed
to recover the pseudoscalar singlet mass from the non-singlet pion.
In \cite{FNAL_00}, the unique piece of the $\eta'$ correlator -- the
hairpin -- was computed directly. The parameter $m_0^2$ was extracted
and one sees the lattice data is well described by the \chiPT\
prediction.  With the shift, the $\eta'$ mass (at non-zero lattice
spacing) is determined to be $820(30)$MeV with possibly large ${\cal
O}(a)$ scaling uncertaintities. A recent Domain Wall
calculation\cite{BNL_02} gives $940(4)$MeV.

\begin{figure}[t]
\mbox{\parbox{0.4\textwidth}{
\parindent=20pt
\epsfig{figure=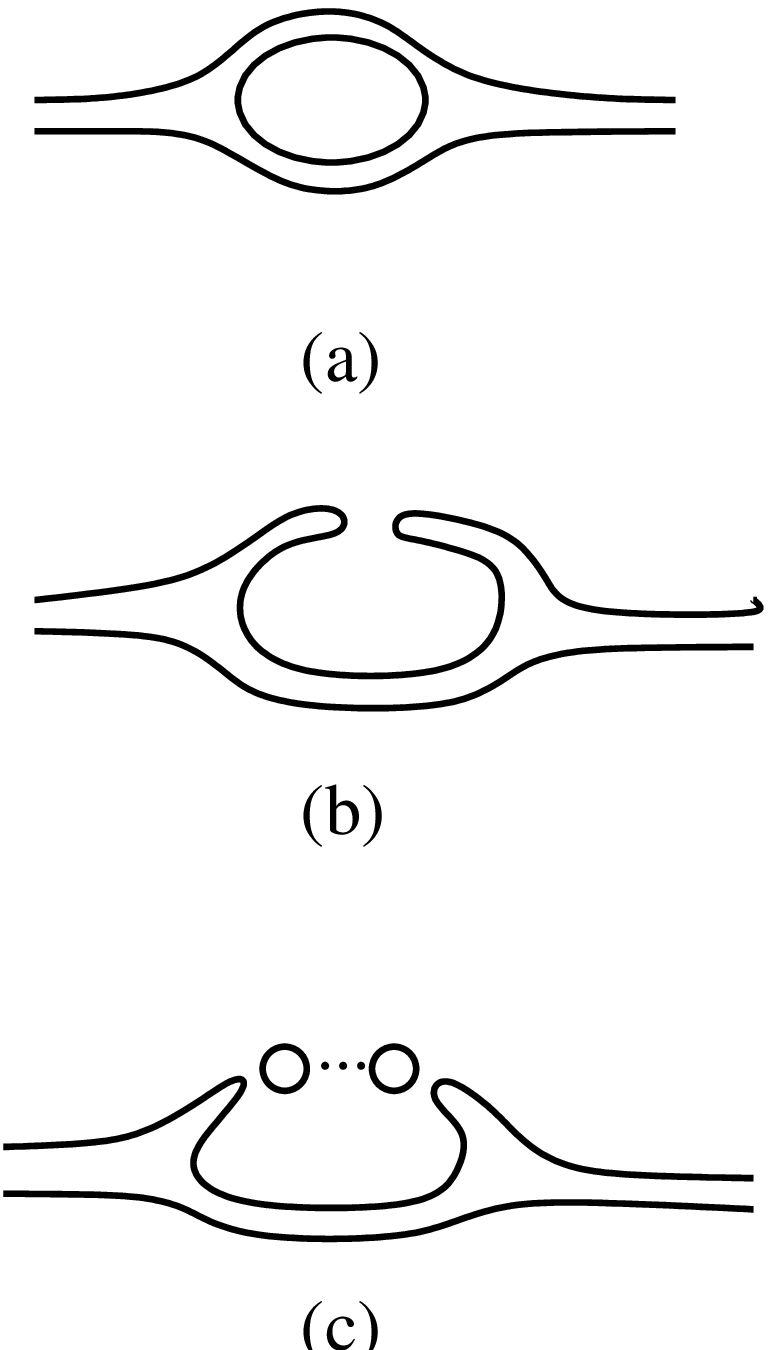, width=2.0cm} 
}} \hfil \mbox{\parindent=0pt
    \parbox{1.0\textwidth}{
    \epsfig{figure=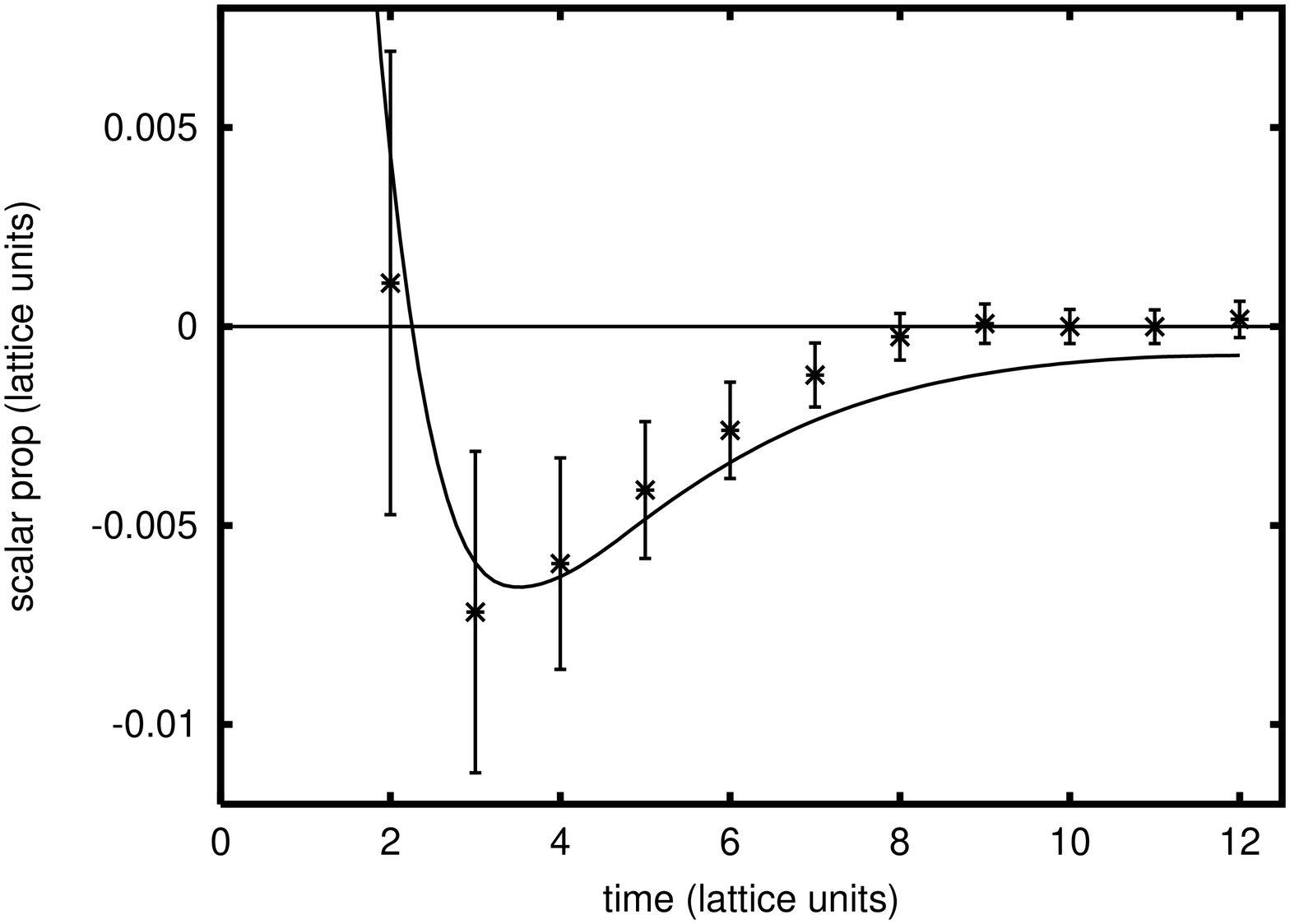, angle=270, width=5.5cm}}}

\caption{Left: contributions to the $a_0$ propagator
from an $\eta-\pi$ intermediate state. Right: 
comparison of scalar $a_0$ propagator with
the bubble sum formula 
fitted to the interval t=1-6
(Ref.\protect\cite{FNAL_01}).}
\label{fig:a0}
\end{figure}

Further dramatic behavior is seen in the isotriplet scalar particle
$a_0$. There is an $\eta'-\pi$ intermediate state with missing
contributions in the quenched approximation as shown in
Fig.~(\ref{fig:a0}). In fact, the $a_0$ correlator goes negative - a
clear sign of violations of unitarity\cite{FNAL_01}. From \chiPT, one
can construct the $a0$ correlator by including couplings between
$\eta'-\pi$ states and rescattering states which can be resummed. The
lightest $a0$ correlator is fairly well described by a 1-loop resummed
bubble term with $\eta'$ mass insertion fixed. A mass $m_{a_0} =
1.34(9)$GeV was found. The new Domain Wall calculation\cite{BNL_02}
gives $m_{a_0} = 1.04(7)$GeV.  The latter results does not exclude the
possibility of $a0(980)$ being a $\bar{q}q$ state.


\subsection{Quenched and Full QCD Hadron Spectrum}

The quenched low-lying hadron spectrum has been extensively studied by
the CPPACS collaboration using Wilson
fermions\cite{CPPACS_99}. Masses were computed at four lattice
spacings and extrapolated to the continuum limit. Lattice sizes ranged
up to $64^3\times 112$ for a $3.2$fm box. At each lattice spacing, the
lightest pseudoscalar mass obtained was about $500$ to $600$MeV. 
Hadron masses were extrapolated in the quark mass via an ans\"atz motivated
by the quenched \chiPT\  prediction
\begin{equation}
\begin{array}{l}
 m_{PS,12}^2 = A(m_1 + m_2)
   \{1 - \delta \left[{\ln \left({2A m_1 /\Lambda_{\rm X}^2}\right)} \right] \\
  \qquad +\  m_2 /\left(m_2 - m_1\right)\ln \left(m_2 /m_1\right) \} 
  + B\left( {m_1  + m_2 } \right)^2 \, + \,{\rm O}\left( {m^3 } \right) \\ 
 m_H \left( {m_{PS} } \right)\, = \,m_0 \, + \,C_{1/2} m_{PS}  + C_1 m_{PS}^2 
  + C_{3/2} m_{PS}^3 \,,\quad C_{1/2}
    \propto \,\delta \ . \\ 
\end{array}
\label{eq:cppacs_fit}
\end{equation}
The constants $A$, $B$, and $C_i$ are fixed in the fits.  Quenching
effects were more clearly seen in the pseudoscalar channel. The
calculation shows the basic hadron spectrum is well determined even in
the quenched approximation to within 10\% accuracy.  The computational
cost was roughly $50$ Gigaflop-years.

A subsequent two-flavor dynamical calculation was made with four quark
masses at 3 lattice spacings\cite{CPPACS_00}. Box sizes range up to
about $2.5$fm. In the meson sector, the results are consistent with
the original quenched calculations and now agree to within 1\% of
experiment.  Systematic deviations of the quenched calculation from
experiment are seen demonstrating that sea quark effects are
important.  The vector meson masses are increased after unquenching.
This increased hyperfine splitting is consistent with the qualitative
view that the spin-spin coupling in quenched QCD is suppressed
compared to full QCD due to a faster running of the coupling constant.

\begin{figure}[t]

\mbox{\parbox{0.4\textwidth}{
\parindent=0pt
\epsfig{figure=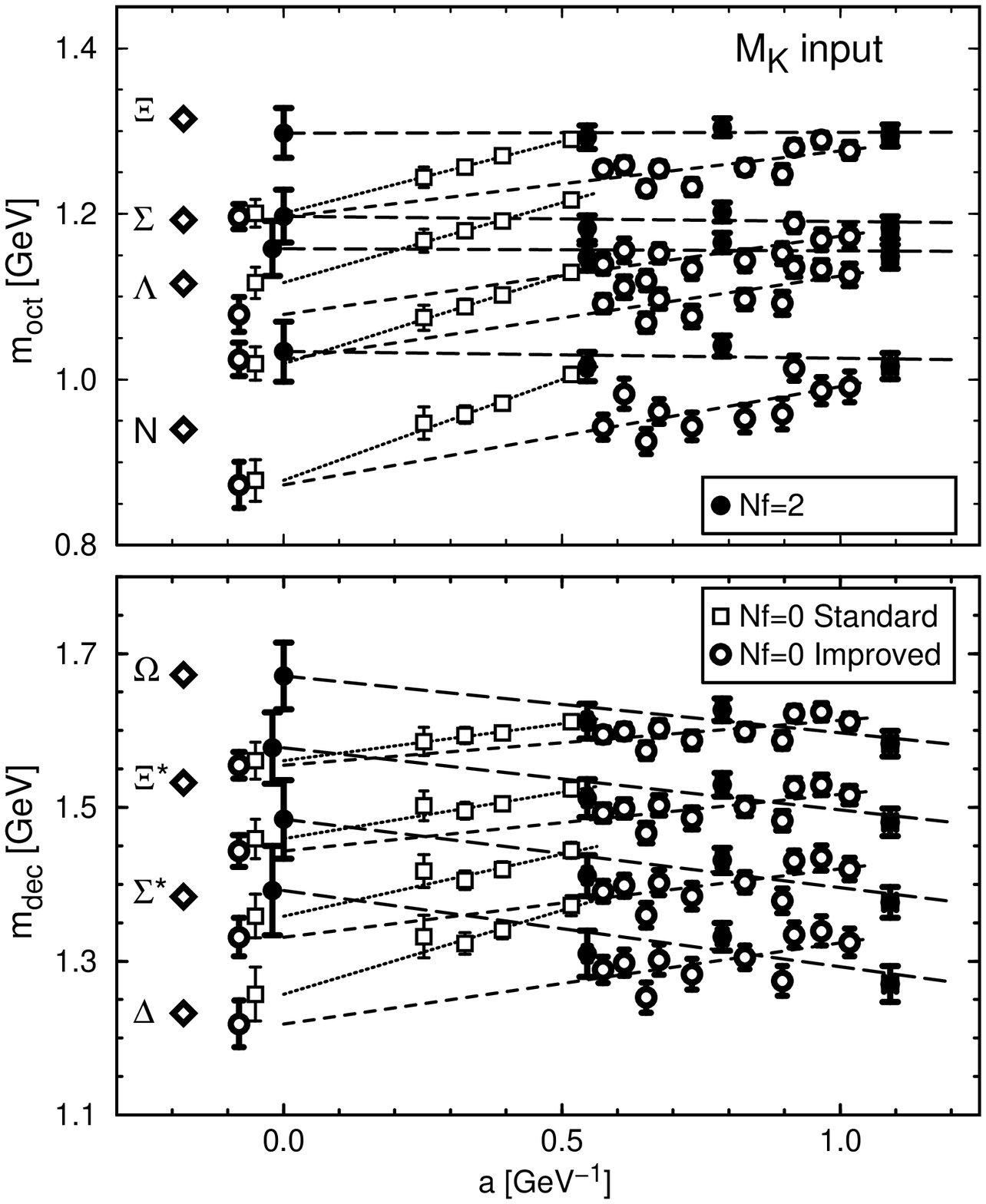, bbllx=48, bblly=82, bburx=550, bbury=385, clip=, width=5.0cm, height=4.5cm} 
}} \hfil \raisebox{5mm}{
    \parbox{0.5\textwidth}{\parindent=0pt
    \epsfig{figure=baryon.Kinp.eps, bbllx=48, bblly=385, bburx=550, bbury=690, clip=, width=5.0cm, height=4.5cm}}}

\caption{Baryon masses in two-flavor (filled
symbols) and quenched (open symbols) QCD. Both graphs have the same lattice spacing
scale.
(Ref.\protect\cite{CPPACS_00}).}
\label{fig:ContBar}
\end{figure}

In the baryon sector, two-flavor dynamical sea quark effects are not
as apparent.  As seen in Fig.~(\ref{fig:ContBar}). The $N$ and
$\Delta$ masses are higher than experiment, but other masses are
consistent. With only a $2.5$fm box, finite-volume effects could well
be large. Another concern is that the octet and decuplet chiral
extrapolations have many parameters resulting in possibly
underestimated errors and will be discussed more next. However, this
is a significant calculation involving roughly one Teraflop-year
of computations and well demonstrates the efficacy of lattice methods.

\subsection{Improved Chiral Extrapolations}
The Adelaide group has been extensively studying higher order \chiPT\ 
effects on hadronic quantities\cite{Adelaide_01}. The basic upshot is
that the na\"ive chiral extrapolations in use are just too na\"ive! In
particular, they incorporate leading non-analytic behavior from heavy
baryon \chiPT\ arising from $B\rightarrow B'\pi\rightarrow B$
intermediate states with $B=N,\Delta$
\begin{equation}
M_B  = \alpha_B  + \beta_B m_\pi^2  + \Sigma_B \left({m_\pi ,\Lambda}\right)
\label{eq:bar_sig}
\end{equation}
where $\Sigma_B$ is a self-energy term. The coefficient of the
$m_\pi^3$ term is actually known analytically in contrast to
Eq.~(\ref{eq:cppacs_fit}).  The basic argument is since \chiPT\ has a
zero radius of convergence (or certainly not a well defined radius), a
simple leading order approximation to Eq.~(\ref{eq:bar_sig}) is quite
a bad approximation at moderate $m_\pi$.  They use a simple
regularization of the self-energy term.

\begin{figure}[t]
\mbox{\parbox{0.4\textwidth}{
\epsfig{file=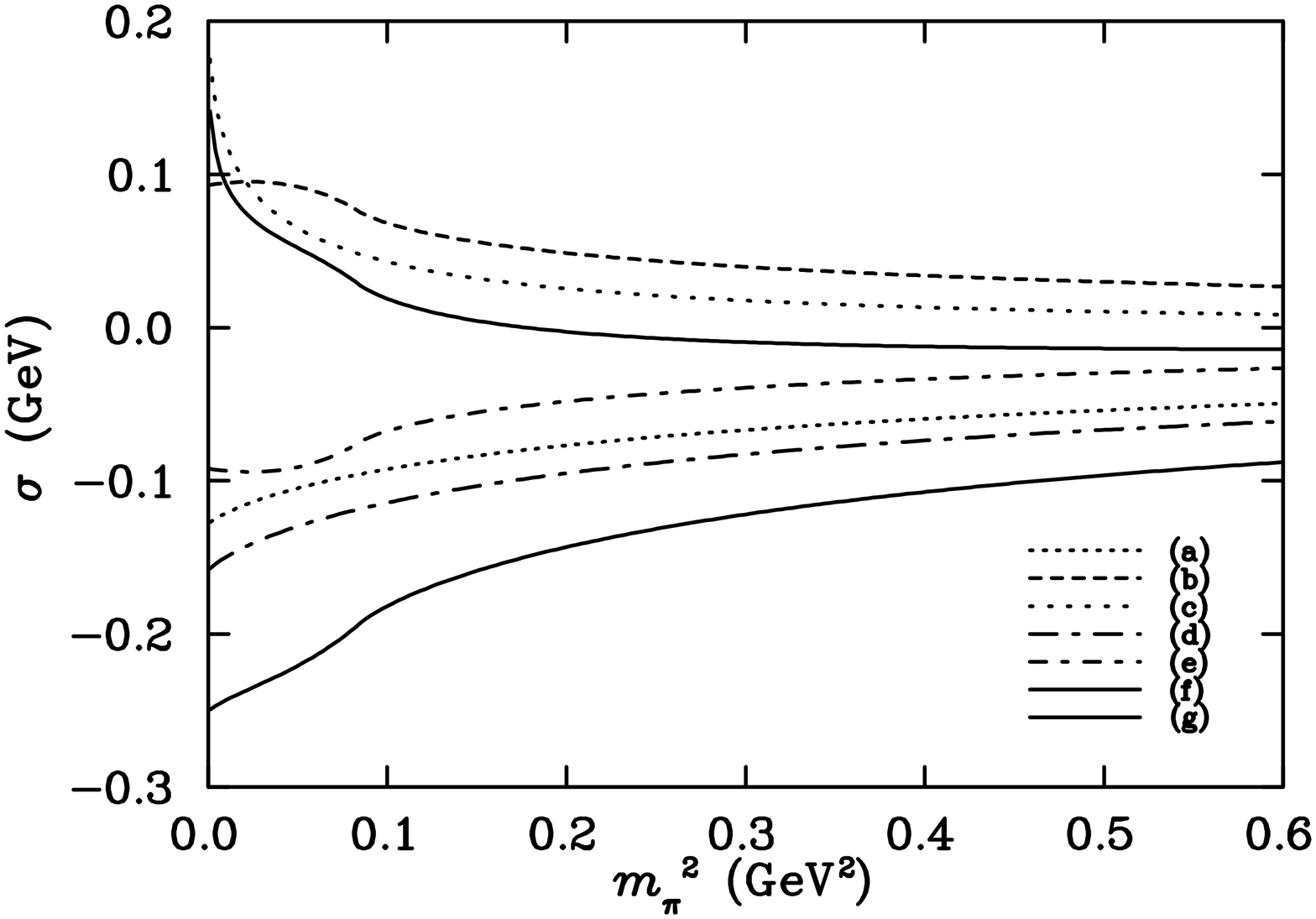, width=4.0cm, angle=90}
}} \hfil \mbox{\parindent=0pt
    \parbox{0.6\textwidth}{
    \null\hfill\epsfig{figure=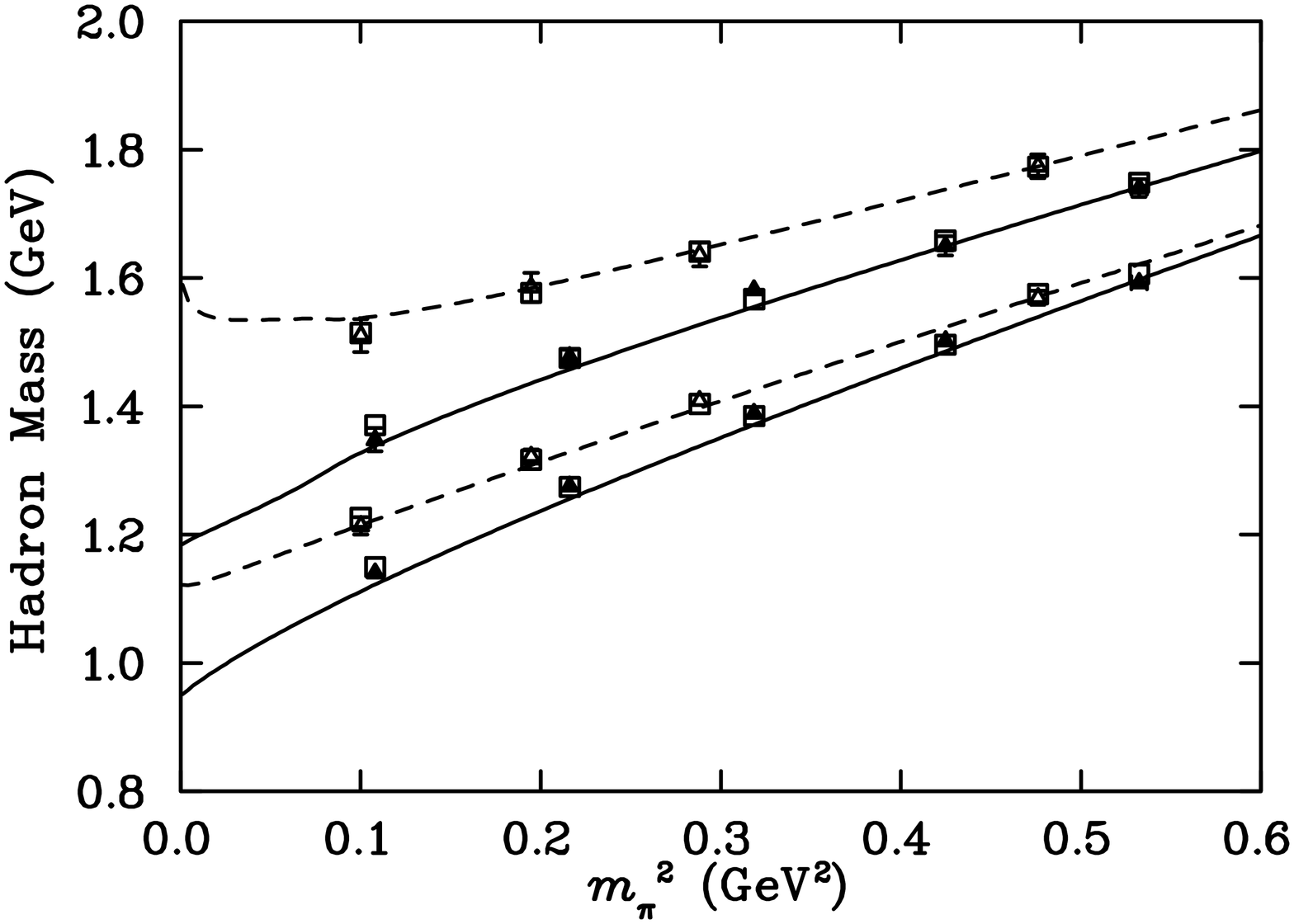, clip=, width=4.0cm, angle=90}}}

\caption{
Left: contributions from various intermediate states to the
quenched and unquenched self-energy term $\Sigma$. Right: fit (open
squares) to lattice data - quenched (open $\triangle$) and dynamical
(filled $\triangle$) with adjusted self-energy expressions accounting
for finite volume and lattice spacing artifacts.  The continuum limit
of quenched (dashed lines) and dynamical (solid lines) are shown. The
lower curves are for $N$ and upper for $\Delta$.
(Ref.~\protect\cite{Adelaide_01}.)}
\label{fig:fqFit}
\end{figure}

Fig.~(\ref{fig:fqFit}) shows a comparison of a recent quenched
and $2+1$ dynamical calculation of the low lying hadron spectrum for a
variant of staggered fermions at the same (physical) lattice
spacing\cite{Adelaide_01}.  The various lines are from fits using
Eq.~(\ref{eq:bar_sig}).  One can see that the self energy term becomes
significant for small lattice quark masses. The intriguing result is
that the fit parameters $\alpha_B$ and $\beta_B$ agree very well
between the quenched and dynamical calculations. This result can be
used to justify the claim that the dominant effects of quenching is
attributed to first order meson loop corrections.

While quite intriguing it is fair to say there is some
controversy over these results. At issue is the concern that once one uses any
model to directly interpret lattice results, one has lost predictably. 
However, in defense once one used a chiral extrapolation at all one has
chosen a model. Ultimately, the Adelaide's group work has shown that
their is interesting structure in the ``pion cloud'' around a hadron and
going to light quark masses is essential.

\subsection{Excited Baryons}

Understanding the $N^*$ spectrum gives vital clues about the dynamics
of QCD and hadronic physics. Some open mysteries are what is the
nature of the Roper resonance? Why is the ordering of the
lowest-lying states - the positive and negative parity states -
inverted between the $N$, $\Delta$ and $\Lambda$ channels?

The history of lattice studies of excited baryons is quite
brief. Recently, new calculations are starting to appear using
improved gauge and fermion actions. The nucleon channel is the most
studied and work has focussed on two independent local interpolating
fields
\begin{equation}
 N_1 = \varepsilon_{ijk} \left( {u_i^T C\gamma_5 d_j} \right)u_k\ , \qquad
 N_2 = \varepsilon_{ijk} \left( {u_i^T Cd_j } \right)\gamma_5 u_k \ .
\label{eq:interp}
\end{equation}
Both interpolating fields couple to positive and negative parity
states, so in practice parity projection techniques are used. Making
the lattice anisotropic with finer discretization in time allows the
behavior of the correlators to be examined over many more time slices
than on isotropic lattices. Additional tuning of the fermion action is
needed to recover hypercubic symmetry. 

\begin{figure}[t]
\mbox{\parbox{0.5\textwidth}{
\epsfig{figure=aniso2.ps, clip=, width=5.5cm}
}} \hfil \mbox{\parindent=0pt
    \parbox{0.5\textwidth}{
    \epsfig{figure=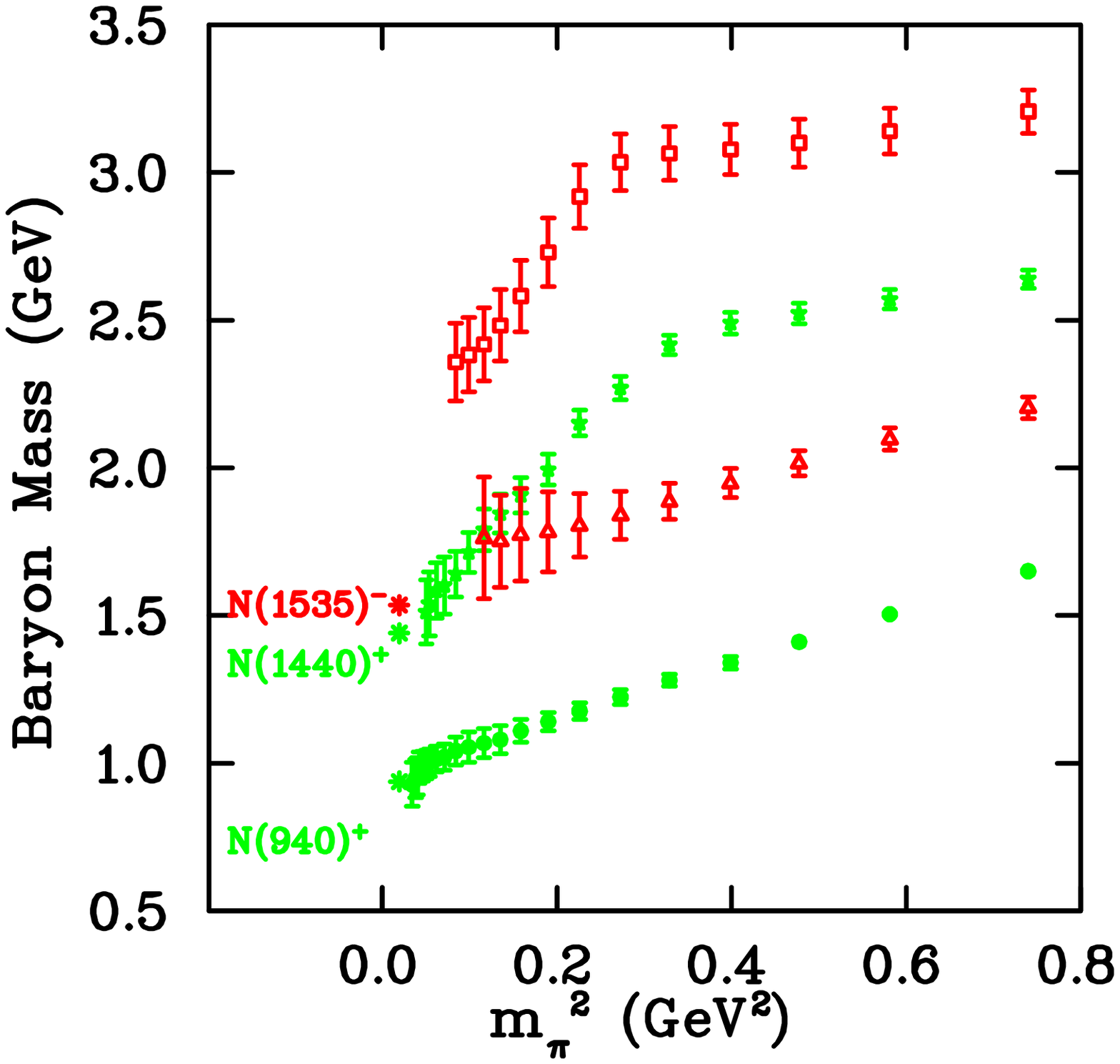, clip=, width=4.5cm}}}
\caption{Left: effective masses for correlators with fields Eq.~(\protect\ref{eq:interp})
corresponding to the nucleon (circles), its parity partner (diamonds) and tentatively
the lowest positive parity excitation (cross) with an anisotropic clover action
(Ref.\protect\cite{Edwards_02}).
Right: masses in physical units obtained with the isotropic Overlap action
(Ref.\protect\cite{Lee_02}).
Solid symbols denote $N(\frac{1}{2}^+)$ states: 
ground ($\bullet$) and 1st-excited ($\star$).
Empty symbols denote $N(\frac{1}{2}^-)$ states: 
lowest ($\triangle$) and 2nd lowest (\fbox{}).
The experimental points ($*$) are taken from PDG.}
\label{mass_nuc_dong}
\label{fig:nstar}
\end{figure}

The left side of Fig.~(\ref{fig:nstar}) shows the effective mass for
the nucleon and its parity partner on an anisotropic lattice using the
clover action\cite{Edwards_02} -- an action improved to have reduced
discretization uncertainties. Long plateaus (clean extraction of a
mass) are seen demonstrating the efficacy of the method. For this
quark mass, there appears to be the expected ordering of the states
$N^{* 1/2+} > N^{1/2-} > N^{1/2+}$. However, there is difficulty in
approaching small quark masses, and the mass extracted with the
$N_2$ operator appears too large. 

A recent quenched calculation\cite{Lee_02} using Overlap fermions (a
chiral fermion action) going to much smaller quark masses reveals a
dramatic decrease in the nucleon masses extracted only with the $N_1$
interpolating field as seen in the right side of
Fig.~(\ref{fig:nstar}). The authors claim the apparent crossing of the
first excited $N^{1/2}$ and lowest $N^{1/2-}$ states is the
demonstration of the physically correct ordering of states. However,
it is also possible that at such light quark masses a decay threshold
has been crossed and the mass observed is affected by missing
dynamical effects via the mechanisms described in
Sec.\ref{sec:decay}. If so, a finite volume check via shrinking the
lattice box can reveal this. It is clear lattice calculations are
really beginning to probe interesting excited state phenomena. With
judicious use of finite volume techniques, physically relevant mass
information can be extracted.

\section{Conclusions}

First generation lattice calculations of excited baryon spectroscopy
are appearing.  
State of the art calculations
require roughly 100 Gigaflop-year in quenched QCD and roughly 1 to 10
Teraflop-years in full QCD. The required resources are not available
to the US lattice community. The Dept. of Energy's SciDAC program is
addressing this shortcoming and a large effort is ongoing in the
U.S. to meet future computational needs.

\section*{Acknowledgments}

RGE was supported by DOE contract DE-AC05-84ER40150 under which the
Southeastern Universities Research Association (SURA) operates the
Thomas Jefferson National Accelerator Facility (TJNAF).


\begin{thebibliography}{99}

\bibitem{Montvay_book} I. Montvay and G. M\"unster, 
{\em Quantum Fields on a Lattice},
(Cambridge Univ. Press, 1994).

\bibitem{Kharzeev_96} 
D. Kharzeev, \Journal{\PLB}{378}{238}{1996}.

\bibitem{Isgur_paton} 
N.~Isgur and J.~Paton,
\Journal{\PRD}{31}{2910}{1985}.

\bibitem{Cornwal_XX} 
J.~M.~Cornwall,
\Journal{\PRD}{54}{6527}{1996}.

\bibitem{Takahashi_02}
T.~T.~Takahashi, {\em et.al.},
\Journal{\PRD}{65}{114509}{2002}.

\bibitem{Alexandrou_02} 
C.~Alexandrou, P.~De Forcrand and A.~Tsapalis, hep-lat/0209062.

\bibitem{Simonov_02}
D.S. Kuzmenko, Yu.A. Simonov, hep-ph/0202277.

\bibitem{Chiral} D.B. Kaplan, Phys. Lett. {\bf B288}, 342 (1992).
R. Narayanan and H. Neuberger, Nucl. Phys. {\bf B443}, 305 (1995).

\bibitem{Edwards_98} 
R.G.~Edwards, U.M~Heller, T.R.~Klassen, \Journal{\PRL}{80}{3448}{1998}.

\bibitem{FNAL_00} B. Bardeen, {\em et.al.}, \Journal{\PRD}{62}{114505}{2000}.
\bibitem{chipt} C. Bernard and M. Golterman, \Journal{\PRD}{46}{853}{1992}.
  S. Sharpe, \Journal{\PRD}{46}{3146}{1992}.
\bibitem{BNL_02} S. Prelovsek and K. Orginos, hep-lat/0209132.
\bibitem{FNAL_01} B. Bardeen, {\em et.al.}, \Journal{\PRD}{65}{014509}{2002}. 
\bibitem{CPPACS_99} S. Aoki, {\em et.al.}, \Journal{\PRL}{84}{238}{2000}.
\bibitem{CPPACS_00} A. Ali Khan, {\em et.al.}, \Journal{\PRD}{65}{054505}{2002}.
\bibitem{Adelaide_01} R.D. Young, {\em et.al.},
hep-lat/0111041.
\bibitem{Edwards_02} R.G. Edwards, U. Heller, D. Richards, proceedings of Lattice 02.
\bibitem{Lee_02} F.X. Lee, {\em et.al.}, hep-lat/0208070.


\end{thebibliography}
\end{document}